\documentstyle[aps,psfig]{revtex}
\newcommand{\eq}[1]{(\ref{eq:#1})}

\begin{document}
\title{ Driving the atom by atomic fluorescence: analytic results for
the power and noise spectra }
\author{Roberta Marani}
\address{Institut d'Optique, BP 147, 91403 Orsay, France.}
\author{Maurizio Artoni}
\address{Department of Physics, University of Essex, Colchester CO4
3SQ, England.}
\date{\today}
\maketitle

\begin{abstract}
We study how the spectral properties of resonance fluorescence propagate
through a
two--atom system.
Within the weak--driving--field approximation we find that, as we go from one
atom to the next, the power spectrum exhibits both sub--natural linewidth
narrowing
and large asymmetries while the noise spectrum of the squeezed
quadrature
 narrows but remains
otherwise
unchanged.
Analytical results for the observed spectral features of the fluorescence
are provided and their origin is thoroughly discussed.
\end{abstract}

\section{Introduction}
\label{sec:intro}

Since the prediction \cite{mollow:69} and observation \cite{shud-etal:74}
of the three--peaked fluorescence spectrum of a strongly driven two--level
atom, the spectral features of atomic fluorescence have both provided
fundamental insight into the photon--atom interaction and revealed a
variety of
interesting quantum features. Photon anti--bunching together with
sub--Poissonian counting statistics have long been predicted
\cite{carm-etal:76,kimb-mand:76,mandel:79} and experimentally observed
\cite{kimb-mand:77prl,cres-etal:82,shor-mand:83}.
The phenomenon of squeezing in resonance fluorescence has also been
the object of a rather intense research activity \cite{loudon:84,coll-etal:84}.
Most of this work has concerned itself with the non--classical features
of fluorescence from an atom excited by a {\em classical light} source,
typically a coherent light beam.
Over the past decade much interest has also been devoted to the issue
of an atom driven by a {\em non--classical light} source.
In the pioneering work of Gardiner~\cite{gardiner:86},
Carmichael~\cite{carm-etal:87}, Zoller~\cite{ritc-zoll:88}
and their co--workers this was studied by considering a
model of a two--level atom interacting with a broad-band squeezed vacuum.
In spite of its simplicity, such a model has brought forward quite
a number of interesting predictions, of which the inhibition of the
atomic dipole decay due to the reduced quantum fluctuations of the
squeezed vacuum reservoir is perhaps the most ubiquitous one.
These early results have stimulated much interest \cite{reviews} on the
issue of
exciting atomic systems with more realistic forms of non--classical light
such as narrow bandwidth
squeezed light
 \cite{vyas-sing:92}, anti--bunched \cite{gardiner:93} and thermal
\cite{hild-etal:83,gard-park:94}
 light, and the fluorescence from a high--{\em Q} atomic
cavity \cite{rice-carm:88}. Amongst the latest successes in the
field, it is certainly worth mentioning a landmark experiment on
trapped Cesium atoms carried out by Kimble and co--workers \cite{geor-etal:95};
they showed that the two--photon atomic excitation rate,
which is typically quadratic with the intensity of the exciting
field, may become instead linear for an exciting squeezed light
field as long anticipated by the theory \cite{geabanacl:89,java-goul:90}.

Along this line we address the issue of how spectral properties of
atomic fluorescence are modified by an atomic medium where succeeding atoms
are driven only by the output of the previous one with no feedback. 
Here, in particular, we study the case of two atoms one of which is driven by a coherent 
light beam. We are interested in the power and quadrature noise spectra for both
on-- and off--resonance excitation and for which we give analytical 
expressions.
The physics of the off-resonant case is much more complex than the resonant one
and has not been discussed much in the literature.
Our work complements other recent investigations
\cite{gardiner:93,gard-park:94,koch-carm:94} in which the idea of one atom
driven by the fluorescence from another atom is examined.
These works  consider only on--resonance excitation and look at different
features of the output light.

Since non--classical effects in atomic fluorescence usually take
place for weak--field excitations we calculate the power and
quadrature noise (squeezing) spectrum of the fluorescence emitted by each of the two
atoms in the weak--field limit.
The atomic correlations functions needed to obtain such spectra
have recently been derived in \cite{arto-mara:96};
the main expressions for the correlation functions
are thus briefly stated in Sec.~\ref{sec:I} while the expressions for the
power and noise spectral density of the fluorescent light are
explicitly derived in Sec.~\ref{sec:II} and \ref{sec:III}.
The physical interpretation of these results is also given in these
last two sections while a summary and conclusions are given in
Sec.~\ref{sec:V}.

\section{Correlation functions}
\label{sec:I}

Consider a collection of identical two--state atoms, whose energy
levels are separated by $\hbar\omega_0$, interacting with the
radiation field via dipole interaction.
Let $\sigma_{-,n}(t)$ be the lowering operator for the ${n}$th atomic
dipole, $R_{n}(t)$ be the population inversion operator and
$\bbox{\cal A}^{(+)}({\bf r}_n,t)$ be the positive frequency part of
the vector potential of the electromagnetic field at the position of
the ${n}$th dipole. In terms of the slowly--varying dynamical
variables $b_n(t)\equiv \sigma_{-,n}(t) \exp(i\omega_0 t)$ and
$A^{(+)}_n(t)\equiv \bbox{\mu}_n\cdot \bbox{\cal A}^{(+)}({\bf
r}_n,t) \exp(i\omega_0 t)$,
where $\bbox{\mu}_n$ is the induced electric dipole moment, the Heisenberg
equations of motion for the atomic operator (in the rotating wave
approximation) read as \cite{vyas-sing:92}

\begin{eqnarray}
\label{eq:b-n}
d b_n/dt &=& -\beta b_n + 2\omega_0\hbar^{-1} R_n A^{(+)}_n(t)\\ 
d R_n/dt &=&
-2\beta(R_n+1/2)-\omega_0\hbar^{-1}[b^{\dagger}_nA^{(+)}_n(t) +
A^{(-)}(t)b_n]
\end{eqnarray}
We denote here by $\beta$ half the natural radiative decay rate.
After the ${n}$th dipole has
been excited, the far--field scattered radiation is
described by the vector
potential (C.G.S. units)

\begin{equation}
\label{eq:A}
{\bf A}^{(+)}({\bf r},t) = \frac{-i
\bbox{\mu}_n\omega_0}{6\pi\epsilon_0 c^2
|{\bf r} - {\bf r}_n|} b_n(t-|{\bf r} - {\bf r}_n|/c) + {\bf
A}^{(+)}_{free}({\bf r},t)
\end{equation}
The first term is the well--known expression for the field radiated
by an oscillating
 dipole located at ${\bf r}_n$, while the second term is the
source--free field contribution
 that has excited the dipole.

The degree of first and second--order coherence needed to evaluate the
spectrum of the resonance
fluorescence and the quadrature noise spectrum 
can be derived from Eq.(\ref{eq:A})
using the solutions of the coupled differential equations (1-2) for
each atomic dipole in
the collection. In particular, the system of Eqs.(1-2) can be transformed
into two coupled integral equations which can be solved, in the weak-field
limit, by a
perturbative technique when
the external driving field only couples to the first atom
and each succeeding atom of
the collection is driven only by the output of the previous one, with
no feedback.

This approach, which is discussed in detail in \cite{arto-mara:96},
is applied here to a system of {\em two atoms}; the second atom is placed
outside
the field which drives the first one and the fluorescence feeding back onto
the
first atom can be clearly assumed to be very small when compared to that
driving
the second atom. When the field impinging on the first atom is a single--mode
{\em coherent state} the steady--state atomic correlation functions,
for the first and second atom are (for positive $\tau$)

\begin{eqnarray}
<b_1(t)> &=& -\frac{\Omega}{2\theta_{-}}\left[1- s_0\right]
\exp(i\Delta t+i\theta_p)\\
<b^{\dagger}_1(t)b_1(t+\tau)> &=& \frac{s_0}{2}
\left[1- 2 s_0\left(1+ \frac{\theta_+ e^{-\theta_-\tau}-\theta_-
e^{-\theta_+\tau} }{4i\Delta}\right)\right] \exp(i\Delta\tau) \\
<b^{\dagger}_1(t) b^{\dagger}_1(t+\tau)> &=&
\frac{s_0}{2}\exp(-2i\phi_+)
\left(
1- e^{-\theta_+\tau}-
\right. \nonumber \\
&&
\left. 2 s_0
\left[
1-e^{-\theta_+\tau} \left(1+\frac{\theta_+\tau}{2}\right)
\right]
\right) \exp(-i\Delta(2t+\tau)-2i\theta_p).
\label{eq:bdagbdag1}
\end{eqnarray}

and

\begin{eqnarray}
<b_2(t)> &=& -\frac{i\Omega\xi_2}{2\theta_-}
\left[1 - s_0 (1+2 s_1) \right]
\exp(i\Delta t +i\theta_p)\exp(-i\Delta|{\bf r_2}-{\bf r_1}|/c) \\
<b^{\dagger}_2(t)b_2(t+\tau)> &=& s_0 s_1
\left\{ 1- 2 s_0
\left[ 1+ 2 s_1
\left( 1+ \frac{\theta_+ e^{-\theta_-\tau}-\theta_- e^{-\theta_+\tau}
}{4i\Delta} \right)
\right] +
\right. \nonumber \\
&&
\left. \frac{s_0}{2(\beta^2+\Delta^2)}
\left[-\frac{\theta_-^2 e^{-\theta_+\tau}}{4\Delta^2} +
e^{-\theta_-\tau}
\left[-\frac{\theta_+^2(\theta_+^2-7\theta_-^2+2(\beta^2+\Delta^2))}
{16\beta^2\Delta^2} -
\right.
\right.
\right. \nonumber \\
&&
\left.
\left.
\left. \frac{2\theta_+^2\theta_-\tau} {4i\beta\Delta} \right]
\right]
\right\}\exp(i\Delta\tau)\\
<b^{\dagger}_2(t)b^{\dagger}_2(t+\tau)> &=&
- s_0 s_1 \exp(-4i\phi_+)
\left\{(1+4 s_0 s_1) \left[ 1-
e^{-\theta_+\tau}\left(1+\frac{\theta_+\tau}{2}\right)\right] -
\right. \nonumber \\
&&\left. 2 s_0\left[1-
\frac{e^{-\theta_+\tau}}{8}(8+5\theta_+\tau+\theta_+^2\tau^2)\right]\right\}
\times \nonumber \\
&& \exp(-i\Delta(2t+\tau)-2i\theta_p)\exp(2i\Delta|{\bf r}_2-{\bf r}_1|/c).
\label{eq:bdagbdag2}
\end{eqnarray}
Here $\theta_{\pm}\equiv \beta \mp i\Delta$,
$\Delta\equiv\omega_0-\omega_L$ is the
detuning between the atomic transition frequency $\omega_0$ and the
frequency $\omega_L$
of the incident laser light, $\phi_+\equiv \arg(\theta_+)$, while $\Omega$
and $\theta_p$ are the Rabi frequency and phase of the
incident field.
 The parameter $\xi_2\equiv (\bbox{\mu}_2\cdot\bbox{\mu}_1
\omega_0^2)/(6\pi\epsilon_0\hbar c^2|{\bf r}_2-{\bf r}_1|)$
represents the strength of the interaction
 between the two dipoles and depends on the direction of propagation
 of the external pump beam with respect to the inter-atomic axis and on
 the inter-atomic distance $|{\bf r}_2-{\bf r}_1|$. For typical parameters of
 experimental interest $\xi_2$ is much less
 than $\beta$.
We have further introduced the saturation parameters
\begin{equation}
s_0\equiv\frac{\Omega^2}{2(\beta^2+\Delta^2)}
\;\;\mbox{\rm and}\;
s_1\equiv\frac{\xi_2^2}{2(\beta^2+\Delta^2)},
\end{equation}
which are both much smaller than one in the weak -field and
far-field
limit respectively.

\section{Power Spectrum}
\label{sec:II}
The incoherent part of the power spectrum of the emitted fluorescence is
given by the
Fourier transform of
 the two--time correlation function as

 \begin{equation}
\label{eq:SE}
S^E_{n}({\bf r},\omega) = \eta \lim_{t \to\infty}
\int_{-\infty}^{+\infty} d\tau \exp(i \omega\tau) <\bbox{\cal E}^-_{n}({\bf
r},t), \bbox{\cal E}^+_{n}({\bf r},t+\tau) >.
\end{equation}
where $\omega$ represents the scattered frequency,
$\eta$ the overall detection efficiency and $<{\bf A},{\bf B}>$
denotes the correlation $<({\bf A}-<{\bf A}>)\cdot({\bf B}-<{\bf B}>)>$.
By using the source--field expression for the electric field
\begin{equation}
\label{eq:En-}
\bbox{\cal E}^-_{n}({\bf r},t)=\bbox{\epsilon}_n({\bf r})
b^{\dagger}_n\left(t-\frac{|{\bf r} - {\bf r}_n|}{c}\right)\exp(i \omega_0 t)
+\bbox{\cal E}^-_{\rm free}({\bf r},t),
\end{equation}
where $\bbox{\epsilon}_n({\bf r}) \equiv \omega_0^2({\bf r}-{\bf
r}_n)\times[\bbox{\mu
}\times ({\bf r}-{\bf r}_n)]/(4\pi\epsilon_0 c^2 |{\bf r}-{\bf
r}_n|^3)$ is
the usual far--field  geometrical factor for the dipole radiation,
and with the help of results (4-9), the spectra of the fluorescence
scattered by the first and second atom are, respectively,

\begin{equation}
\label{eq:S1E}
S_1^E({\bf r},\omega)=\left[\eta \frac{|\bbox{\epsilon}_1({\bf
r})|^2}{\beta}\right] s_0
\frac{\Omega^2\beta^2}{[\beta^2+(\Delta-(\omega-\omega_L))^2]
[\beta^2+(\Delta+(\omega-\omega_L))^2]}
\end{equation}
and
\begin{equation}
\label{eq:S2E}
S_2^E({\bf r},\omega)= S_1^E({\bf r},\omega)
\left|\frac{\bbox{\epsilon}_2({\bf
r})}{\bbox{\epsilon}_1({\bf r})}\right|^2
\left[
4 s_1^2 + 2
s_1\frac{(\beta^2+\Delta^2)}{[\beta^2+(\Delta-(\omega-\omega_L))^2]}
\right].
\end{equation}

The spectrum (\ref{eq:S1E}),
which recovers a known result for the fluorescence power spectrum
of a weakly driven two-level atom \cite{kimb-mand:77,mollow:69}, consists
of two
Lorentzian contributions symmetrically displaced with respect to the
driving frequency
$\omega_L$ by an amount $\pm\Delta$.
 The spectrum (\ref{eq:S2E}), proportional to $S_1^E({\bf r},\omega)$
 through the relative coupling strength
 of the two atomic dipoles, comprises a constant term and a Lorentzian
contribution centered at the value $\Delta$ of the detuning.

It is here worth noting that both the coherent and the incoherent
part of the fluorescence scattered off the first atom contribute to the
spectrum $S_2^E({\bf r},\omega)$. 
This spectrum depends, in fact, through its definition (\ref{eq:SE})
on the atomic dipole operator $b_{2}$ whose time evolution is governed by the 
{\em total} field ${\cal A}^{(+)}$ at {\bf r}$_{2}$, namely 
both coherent and incoherent contributions.

The two spectra (\ref{eq:S1E}) and (\ref{eq:S2E}) are integrated over
the whole solid angle, assuming that all emitted light is collected  by a
detector with perfect efficiency ($\eta$=1), normalized to the output
intensity and plotted in Fig.~\ref{fig1} and \ref{fig2} 
for the case of zero and a non-zero detuning $\Delta$.

The finite bandwidth of the two power spectra has here
a significance that is different from its significance in the process of
atomic spontaneous emission into the vacuum.
An excited atom has only a finite amount of stored energy and can
radiate only for a finite time, so that the width of the spectral
density has to be finite in spontaneous emission processes.
However, in our problem we are dealing with atoms that are continuously
excited and the finite size of the
linewidth is due to the quantum fluctuations of the field driving the atom, 
as we will discuss in the following.


\vspace{\baselineskip}

{\em Resonant excitation.}
The role of quantum fluctuations in the fluorescence spectra  
is already evident in the sub-natural linewidth of the light
emitted by the first atom for zero detuning (dashed curve).
This sub-natural narrowing is experimentally \cite{wu-etal:75} well
established and can be attributed to the atomic dipole fluctuations
induced by the coherent exciting field.
The spectrum $S_1^E$ can in fact be decomposed into two Lorentzians
describing the fluctuations of the two quadrature phase amplitudes of the
induced atomic dipole. Since the fluctuations in one of the
quadratures are squeezed, the corresponding Lorentzian gets a negative
sign and the two contributions sum up to yield 
the spectrum (\ref{eq:S1E}) with a sub-natural linewidth
\cite{note:mollow,rice-carm:88,swai-zhou:96}.

The power spectrum of the second atom $S_2^E$, 
which goes like the third power of the inverse frequency square,
falls off more rapidly than $S_1^E$ does (Fig.1).
This produces a narrowing in the line--shape of the light emitted by the 
{\em second} atom, as
compared to that of the {\em first} atom, which could originate either 
from the correlations in the fluorescence incident on the second atom or
from the fact that the second atom produces its own squeezing and 
subsequent narrowing. 
It can be shown, however, that if the second atom were to be driven by 
a classical field having the same spectral profile of the fluorescence scattered off the first atom
(dashed curve) and made up of monochromatic contributions,
thus neglecting the incident field correlations, 
additional narrowing after the second atom would not occur; the resulting linewidth would be larger
\cite{note:bwidth} than that obtained for monochromatic
excitation (dashed curve) suggesting that the power spectrum narrowing after the second atom
is mainly due to the presence of quantum correlations (squeezing) in
the driving fluorescent light.

Let us also note that the spectral narrowing due to the squeezed
incident fluorescence
should be phase--dependent;
this is however not apparent in our case because the squeezing phase
in the field exciting the second atom cannot be varied at will but it
has a fixed relationship with the phase of the field driving the first atom.

\vspace{\baselineskip}

{\em Off--resonant excitation.}
The case of non--zero detuning exhibits a somewhat more complex physics
(Fig.2).
The light emitted by the {\em first} atom
exhibits a symmetric spectrum with maxima at $+$ and $-\Delta$ from the laser frequency
$\omega_L$.
Such a symmetry
originates from the fact that the two Lorentzian contributions in
(\ref{eq:S1E}) are equally displaced
from $\omega_L$ so that for sufficiently large $\Delta$'s these two
components can be
far enough apart to make up a well separated doublet whose central dip
will decrease as the
separation of the two Lorentzians increases. In the limit of
vanishing detunings, this well
separated doublet
merges into a single line centered at resonance.

Physically, the symmetry of the doublet originates from the fact that in the weak--field 
limit the detuned atom, in order to satisfy energy conservation, 
responds to the absorption
of two laser photons by emitting two photons shifted in frequencies by opposite amounts
with respect to $\omega_L$.
In the dressed--atom approach of resonance fluorescence \cite{cohe-reyn:77}
the symmetric doublet spectrum can be seen to originate from
the suppression of the central peak of the fluorescence triplet in the
spontaneous transitions between the (first) atom's dressed levels.

The spectrum of the fluorescence emitted by the {\em second} atom
displays a strong asymmetry in the peak heights in addition
to an effect of linewidth narrowing.
The dominant term in (\ref{eq:S2E}) does not only produce the
narrowing but it is also responsible for the asymmetric double--peak structure 
of the spectrum.
Narrowing of the linewidth is again caused by the squeezing in the
fluorescence driving the second atom in much the same way as for the undetuned atom.

The fluorescence quenching, which gives rise to the asymmetry, can instead be
explained as follows.
Because the spectrum emitted by the first atom comprises the incoherent part
$S_1^E$ and the coherent part 
$S^E_{1,{\rm coh}}(\omega)=[\eta|\epsilon_1|/\beta]s_0(1-2s_0)\beta\delta(\omega-\omega_L)$.
the light impinging on the second atom draws, according to \eq{S1E}, contributions from a 
range of frequencies scattered over the interval $(\omega_0-2\Delta-\beta,\omega_0+\beta)$ with 
a peak at $\omega_L$ (coherent scattering) and 
two side-peaks centered at about $\omega_0-2\Delta$ and $\omega_0$ (incoherent scattering).
The coherently scattered radiation from the first atom is monochromatic and
yields a symmetric contribution much alike that of the first atom driven by the monochromatic
laser beam; this is the very small term ($s_{1}<<1$) on the right hand side of Eq.(\ref{eq:S2E}).
The incoherently scattered radiation from the first atom, on the other hand, gives rise to the
asymmetric and larger term in Eq.(\ref{eq:S2E}) which mainly enhances the resonant peak;
this contribution arises instead from the single-photon absorption--emission processes.
In fact, each incident $\omega$ scatters into a two--peak spectrum having one peak on
resonance and the other at $2\omega-\omega_0$ so that for each $\omega$ the resonant
component of the spectrum is enhanced resulting into a stronger on--resonance peak 
while the off--resonance components are smeared over the spectrum yielding the observed quenching.

\section{Noise Spectrum}
\label{sec:III}
The resonance fluorescence radiated by
a two--level atom driven by coherent light exhibits the phenomenon of
squeezing \cite{wall-zoll:81}.
This effect is typically revealed by using a phase--sensitive homodyned
detection
scheme \cite{arto-mara:96,ou-etal:87}. In such a scheme the atomic
fluorescence is mixed by means of a symmetric beam splitter
with the highly coherent light of a local
oscillator of the same frequency.
Squeezing in the signal beam can be assessed by measuring the
normally ordered and time ordered intensity correlation function

\begin{equation}
\lambda(t_1,t_2)\equiv \frac{<{\cal T}
:I(t_1)I(t_2):>}{<I(t_1)><I(t_2)>}
 - 1,
\end{equation}
where $I$ is the intensity of the mixed field at the detector.

When the local oscillator is made much stronger than the fluorescent
signal, the intensity correlation function for the fluorescence
scattered from the $n$-th atom is given by:

\begin{eqnarray}
\label{eq:lambda}
\lambda_n({\bf r},t,t+\tau,\psi_n)&=& 2
\frac{|\bbox{\epsilon}_{n}({\bf r})|^2}{|\bbox{\epsilon}_{LO}({\bf r})|^2}
{\rm Re}[\exp({i(\omega_{LO}-\omega_0)\tau})
<b^{\dagger}_n(t_{n,{\bf r}}),b_n(t_{n,{\bf r}}+\tau)>+ \nonumber\\
&&\exp(2 i \psi_n)
\exp({-i(\omega_{LO}-\omega_0)(2t+\tau)})
<b^{\dagger}_n(t_{n,{\bf r}}),b^{\dagger}_n(t_{n,{\bf r}}+\tau)>]
\end{eqnarray}
where $t_{n,{\bf r}}\equiv t-\omega_0 |{\bf r}-{\bf r}_n|/c$ is the typical
retardation time,
$\psi_n\equiv \phi_{LO}+(\pi/2)-(\omega_0 |{\bf r}-{\bf r}_n|/c)$ and
$\phi_{LO}$ is the
local oscillator phase. In homodyne detection the local oscillator
frequency $\omega_{LO}$ is equal to the laser frequency $\omega_L$.
In the stationary regime
$\lambda_n({\bf r},t,t+\tau)$ depends only on $\tau$, and its
Fourier transform yields the quadrature noise spectrum

\begin{equation}
\label{eq:SN}
S^N_n({\bf r},\omega,\psi_n)=\eta <I> \int_{-\infty}^{+\infty} d\tau \exp(i
\omega\tau) \lambda_{n}({\bf r},\tau,\psi_n).
\end{equation}
The phase of the local oscillator $\phi_{LO}$ is chosen
so as to yield maximum squeezing for fixed values of the Rabi frequency and
detuning. This occurs when the intensity correlation function
(\ref{eq:lambda}) acquires the highest negative value for $\tau=0$,
so that $\psi_{1}=\phi_+$ and $\psi_{2}=2\phi_+ + (\pi/2)$ for the
first and second atom, respectively, as can be inferred from
Eqs.(\ref{eq:bdagbdag1}) and (\ref{eq:bdagbdag2}).
In general, the contribution from the first term on the right hand side
of (\ref{eq:lambda}) to the noise spectrum (\ref{eq:SN}) can be written as
$[S^E_n({\bf r},\omega+\omega_L)+S^E_n({\bf r},-\omega+\omega_L)]/2$ while
we will
denote by $\tilde S_{n}({\bf r},\omega,\psi_{n})$ the contribution
coming from the last term on the right hand side of (\ref{eq:lambda}).
Since for the first atom the power spectrum is symmetric
(see Eq.(\ref{eq:S1E})) we obtain
\begin{equation}
\label{eq:S1N}
S_1^N({\bf r},\omega,\pm) = S^E_1({\bf r},\omega+\omega_L) \pm
\tilde S_1({\bf r},\omega,\psi_{1}),
\end{equation}
where
the plus (minus) sign in the third argument of the
noise spectrum corresponds to the squeezed (anti-squeezed)
quadrature, and
\begin{eqnarray}
\tilde S_1({\bf r},\omega,\psi_{1})&=&
\left[ \frac{ \eta |\bbox{\epsilon}_{1}({\bf r})|^2}{\beta}\right]
\left[
2 s_0^2 \left[
\frac{\beta^2+\Delta^2-\omega\Delta}
{(\beta^2+(\Delta-\omega)^2)^2}
+\frac{1}{(\beta^2+(\Delta-\omega)^2)}
\right] -\right. \nonumber \\
&&
\left.
2 s_0\frac{1}{(\beta^2+(\Delta-\omega)^2)}\right]+
\{\omega\rightarrow -\omega\}.
\end{eqnarray}

For the second atom we obtain instead

\begin{equation}
\label{eq:S2N}
S_2^N({\bf r},\omega,\pm) =\frac{ S_2^E({\bf r},\omega+\omega_L)+
S_2^E({\bf r},-\omega+\omega_L)}{2} \pm \tilde S_2({\bf r},\omega,\psi_{2}),
\end{equation}
where
\begin{eqnarray}
\tilde S_2({\bf r},\omega,\psi_{2})&=&
\left[\frac{\eta |\bbox{\epsilon}_2({\bf r})|^2}{\beta}\right]
\left[
- 4 s_1^2 s_0^2
\left[
\frac{2(\beta^2+\Delta^2-\omega\Delta)}
{(\beta^2+(\Delta-\omega)^2)^2}
+\frac{1}{(\beta^2+(\Delta-\omega)^2)}
\right] + \right. \nonumber \\
&&
4 s_1 s_0^2 \left[
\frac{\beta^2+\Delta^2-2\omega\Delta}
{(\beta^2+(\Delta-\omega)^2)^3} +
\frac{2\beta^2+6\Delta^2-3\omega\Delta}
{(\beta^2+(\Delta-\omega)^2)^3} + \right. \nonumber \\
&& \left.
 \frac{3}
{2(\beta^2+(\Delta-\omega)^2)}
\right]  - \nonumber \\
&& \left.
 s_1 s_0\left[
\frac{2(\beta^2+\Delta^2-\omega\Delta)}
{(\beta^2+(\Delta-\omega)^2)^2}
+\frac{1}{(\beta^2+(\Delta-\omega)^2)}
\right]\right]
+\{\omega\rightarrow -\omega\}.
\end{eqnarray}

The spectrum (\ref{eq:S1N}) is consistent with the weak--field limit
of a known expression for the noise spectrum
\cite{coll-etal:84,ou-etal:87}, while (\ref{eq:S2N}) is a new result.
In Fig.~\ref{fig3} we plot the spectrum (\ref{eq:S1N}) and (\ref{eq:S2N})
(squeezed quadrature)
for an undetuned atom and suitably normalized to the outgoing intensity
 while the same is done for a detuned atom in Fig.~\ref{fig4}.
The outgoing intensity is given by
$<b_n^{\dagger}(t)b_n(t)>|\epsilon_n({\bf r})|^2$. We
first observe that for both detuned and undetuned atom
the $noise$ $bandwidth$ narrows as the fluorescence
propagates from one atom to the next because in general the atom--radiation
coupling lengthens the characteristic correlation time of the
scattered fluorescence \cite{arto-mara:96}. This is also consistent
with the findings of Gardiner {\em et al} \cite{gard-etal:87} for a
two--level atom excited by squeezed light from a degenerate parametric
amplifier.

Furthermore, one can see that for an $undetuned$ atom maximum squeezing
is seen to arise in a narrow band around $\omega=0$ where it reaches
an absolute value of about 12\% after the first atom, confirming the
result found in \cite{loudon:84,coll-etal:84,ou-etal:87}, but reduces to
$5\times 10^{-3} \% $ after the second atom.
These percent values are obtained multiplying \eq{S1N} and \eq{S2N}
by the appropriate outgoing intensity.
The very small noise reductions after the second atom
are due to the fact that the effective Rabi frequency for the second atom
is about $10^{-2}$ times smaller than that for the first atom.
For a $detuned$ atom, on the other hand, the amount of squeezing
is generally smaller with largest values occurring
over sidebands centered around $\pm\Delta$. For $\Delta=2\beta$ a maximum absolute
squeezing of about 1\% is observed after the first atom but drops again to
$1\times 10^{-4} \%$ after the second one (Fig.~\ref{fig4}).

Finally we see that the fluorescence from the first atom exhibits homogeneous
squeezing (i.e. squeezing is present at all frequencies), while, for
detunings greater than $2\beta$, the fluorescence
 from the second atom is squeezed only for
frequencies between the side minima. This is illustrated in
Fig.~\ref{fig5}, which shows a contour plot of
the noise spectrum $S^N_2({\bf r},\omega,+)$ normalized to the outgoing
intensity
 as a function of frequency and detuning.
 The white area represents positive values (indicating that the light
is not squeezed at the corresponding frequency) while darker colors
indicate increasing negative values.

\section{Conclusions}
\label{sec:V}

In this paper we calculate the effect of propagation through an atomic
medium on the spectral properties of resonance fluorescence.
For a two--atom system, where the first atom is excited either by a resonant
or non--resonant coherent light source and the second one only by the fluorescence scattered
by the first one (no feedback), we provide analytical expressions for 
the fluorescence
power spectrum and for the fluorescence quadrature noise spectrum (squeezing).
Such a simple system offers the advantage of analytical results which
enable one to stress the basic differences between the spectral signatures
of the fluorescence scattered by an atom driven by a classical (first) and
fluorescent (second) light field.
In the weak--field regime we observe narrowing of the linewidth both
in the power and noise spectra
and an appreciable fluorescence quenching
effect after the second atom.

The mechanisms for line--narrowing after the first and after the second
atom are different. The sub--natural linewidth narrowing in the power
spectra after the first atom is described by
Eq.(\ref{eq:S1E}) and originates from the fact that the interaction between the atom
and a coherent laser source produces squeezing of the atomic dipole--fluctuations
in phase with the mean induced dipole moment.
Further narrowing after the second atom, which is described by
the square bracket term on the right hand side of Eq.(\ref{eq:S2E}), results
from the presence of squeezing in the fluorescent radiation that excites it.

The quadrature noise spectrum of the the first and second
atom fluorescence is instead described by the expressions \eq{S1N} 
and \eq{S2N},
and reductions of the noise bandwidth from one atom to the next are due to the
lengthening of correlation time as the fluorescence propagates from one atom to the other.

Unlike the effect of linewidth narrowing, which occurs for both detuned
and undetuned atoms, quenching of the fluorescence takes place only for
non--vanishing detunings after the second atom.
Such a quenching results from the asymmetry introduced
by the second term in the square bracket in Eq.(\ref{eq:S2E})
and it is here interpreted as due to the enhancement of the resonant
component of the finite--bandwidth radiation impinging on the second
atom.

It is worth mentioning at last that for appropriate values of the detuning
the two--atom system that we examine here is remarkably similar to
the one of an atom driven by a bichromatic field with one resonant
and one off--resonant component of the same intensity. The dynamical
and spectral features of such an apparently simple atomic configuration
have been investigated in a surprisingly small number of experiments; yet
such a system is expected to exhibit quite a rich physics that ranges from
parametric \cite{thomann:80} and Rabi sub--harmonic resonances
\cite{papa-etal:96} to
two--photon optical gain and lasing \cite{gaut-etal:92}.
The compact analytical results derived here may then be of help in
understanding
the not so obvious spectral features of the fluorescence from
a two--level
atom under bichromatic and weak excitation \cite{yu-etal:97}
and work in this direction is under way \cite{preparation}.

\section{Acknowledgements}
\label{sec:end}

The authors are grateful to C. Cabrillo, P. Zhou and S. Swain for enlightening
discussions.


\begin{thebibliography}{10}
\bibitem{note:mollow}
Already in ref. \cite{mollow:69} Mollow notes that the form of the spectrum
  (12) (zero detuning) can be attributed to the addition of two Lorentzian
  components, one of which appears with negative weight.
\end{thebibliography}

\newpage

\begin{figure}
\caption{Spectrum of the fluorescence scattered
by the first (dashed line) and by the second (solid line) atom on resonance.
For all curves $\Omega/\beta= 0.3$. For direct comparison the spectra
(12) and (13) have been scaled so as to acquire the value of unit at the
atomic
transition frequency, i.e. at $\omega=\Delta$.}
\label{fig1}
\end{figure}

\begin{figure}
\caption{Spectrum of the fluorescence scattered
by the
first (dashed line) and by the second (solid line) atom for detuning
$\Delta/\beta = 2$. The other parameters and the scaling are the same
as in Fig. 1}
\label{fig2}
\end{figure}

\begin{figure}
\caption{Normalized noise spectrum (squeezed quadrature) 
of the fluorescence scattered
by the first atom (dashed line) and by the second atom (solid
line) for no detuning, $\Omega/\beta=0.3$, $\psi_1=\phi_+$, $\psi_2=2\phi_+
+ (\pi/2)$ and
$\xi_2/\beta=0.02$.}
\label{fig3}
\end{figure}

\begin{figure}
\caption{Normalized noise spectrum (squeezed quadrature)
of the fluorescence scattered
by the first atom (dashed line) and by the second atom (solid
line) for $\Delta/\beta=2$, $\Omega/\beta=0.3$, $\xi_2/\beta=0.02$,
$\psi_1=\phi_+$,
$\psi_2=2\phi_+ + (\pi/2)$.}
\label{fig4}
\end{figure}

\begin{figure}
\caption{ Contour plot of the noise spectrum (squeezed quadrature)
of the second atom
for $\Omega/\beta=0.3$ and
$\xi_2/\beta=0.02$ as a function of frequency and detuning.
We take $\psi_2=2\phi_+ + (\pi/2)$.
The contour lines represent values of the spectrum from -5 to 0.5,
in 0.5 increments. Lighter areas correspond to higher values. The spectra
are normalized to the outgoing intensity.}
\label{fig5}
\end{figure}

\begin{figure}
\centerline{\psfig{figure=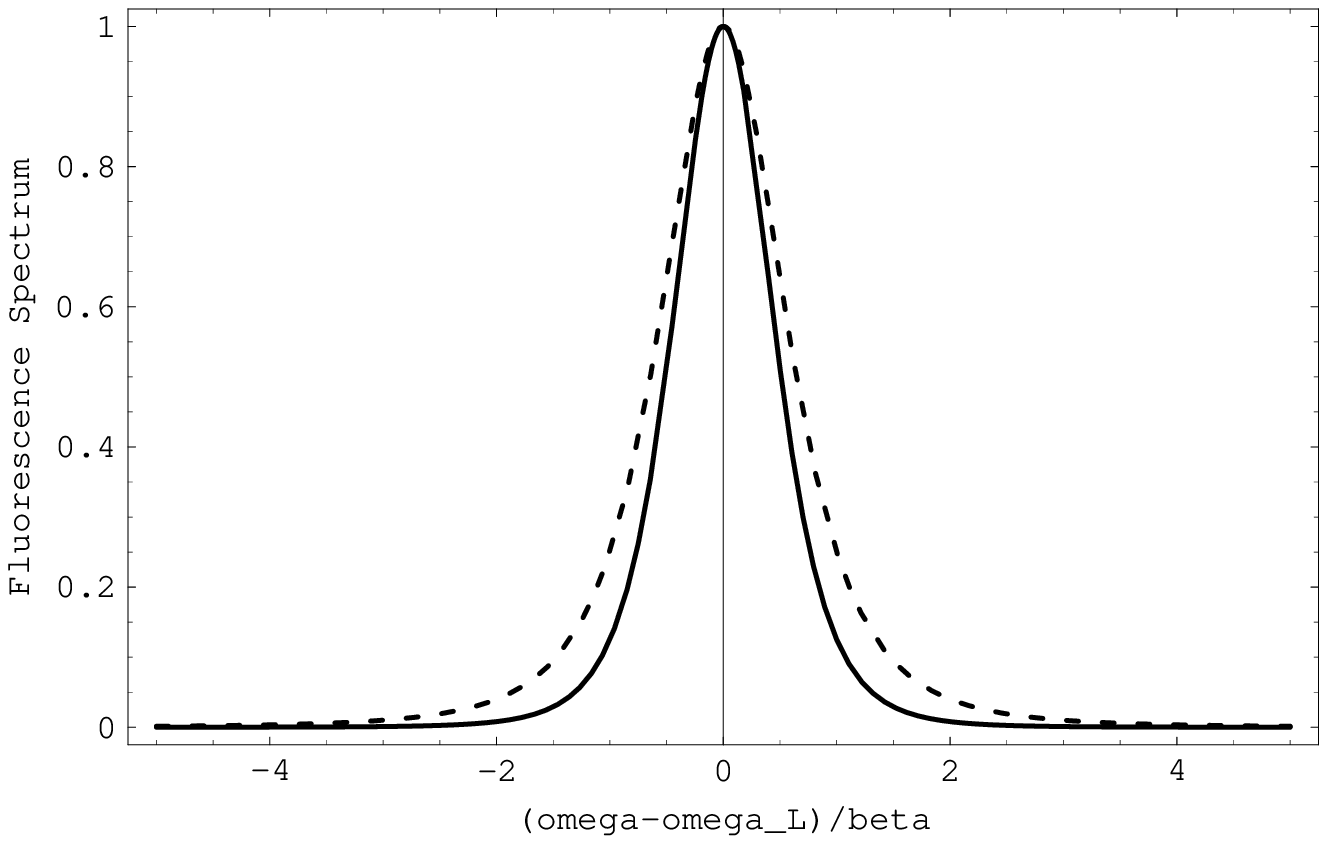}}
\end{figure}

\begin{figure}
\centerline{\psfig{figure=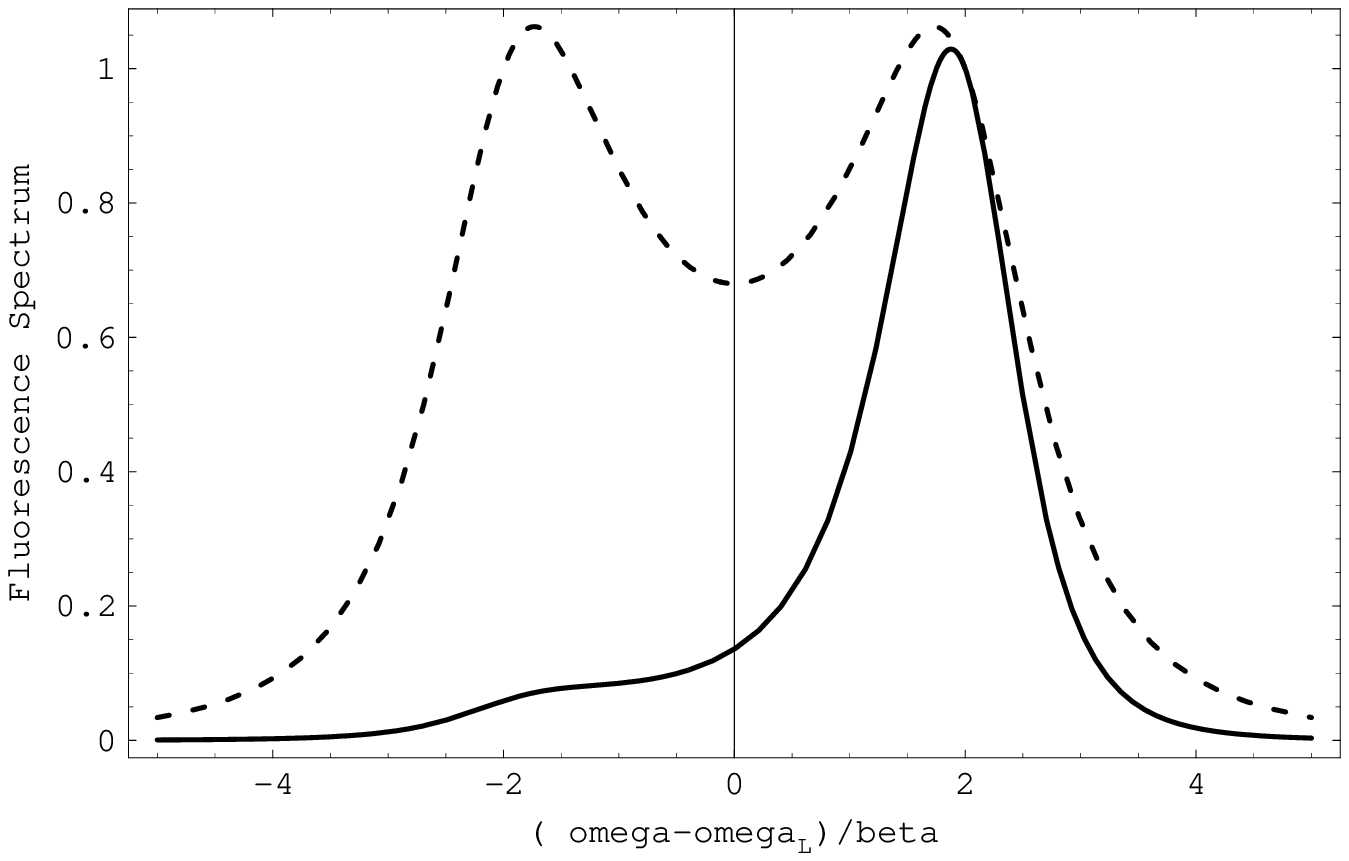}}
\end{figure}

\begin{figure}
\centerline{\psfig{figure=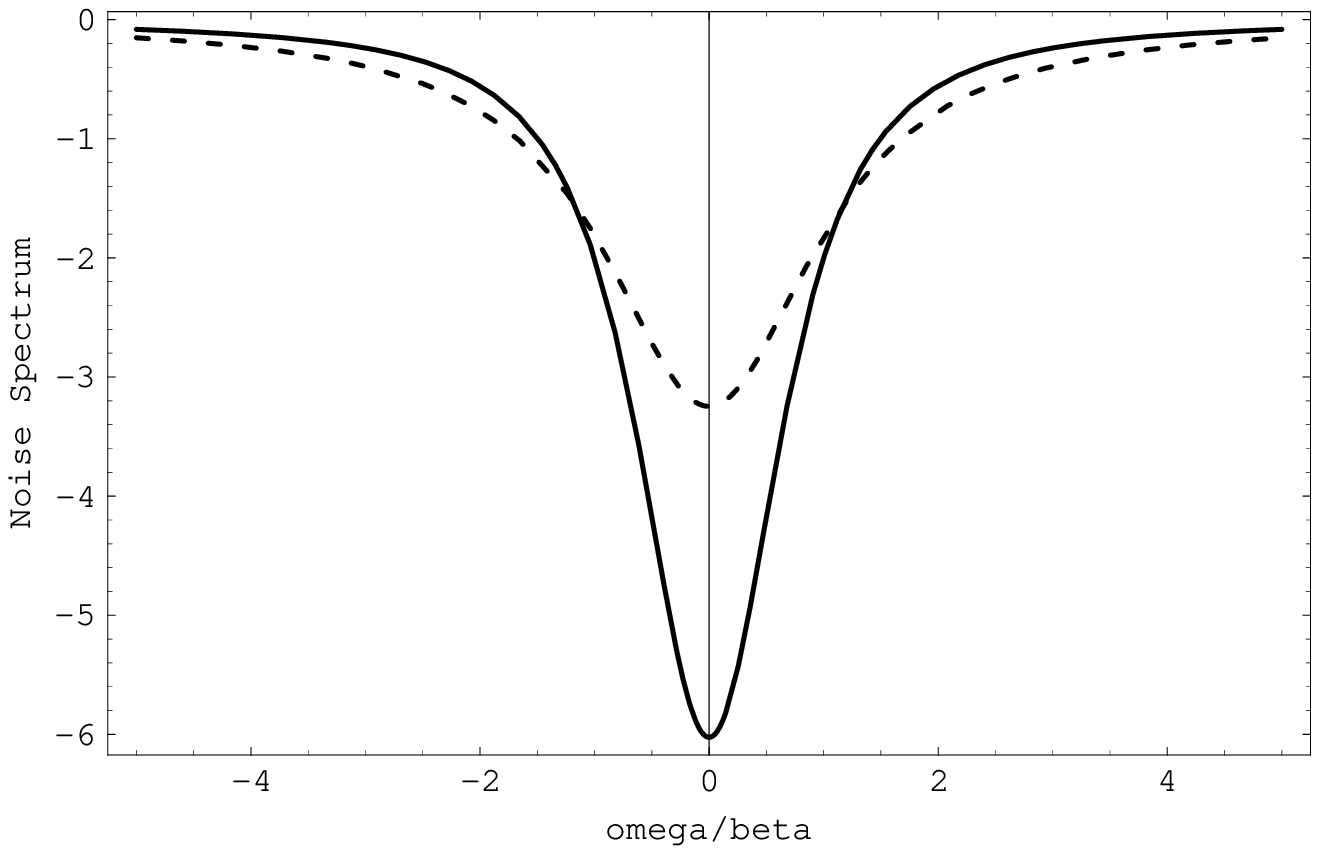}}
\end{figure}

\begin{figure}
\centerline{\psfig{figure=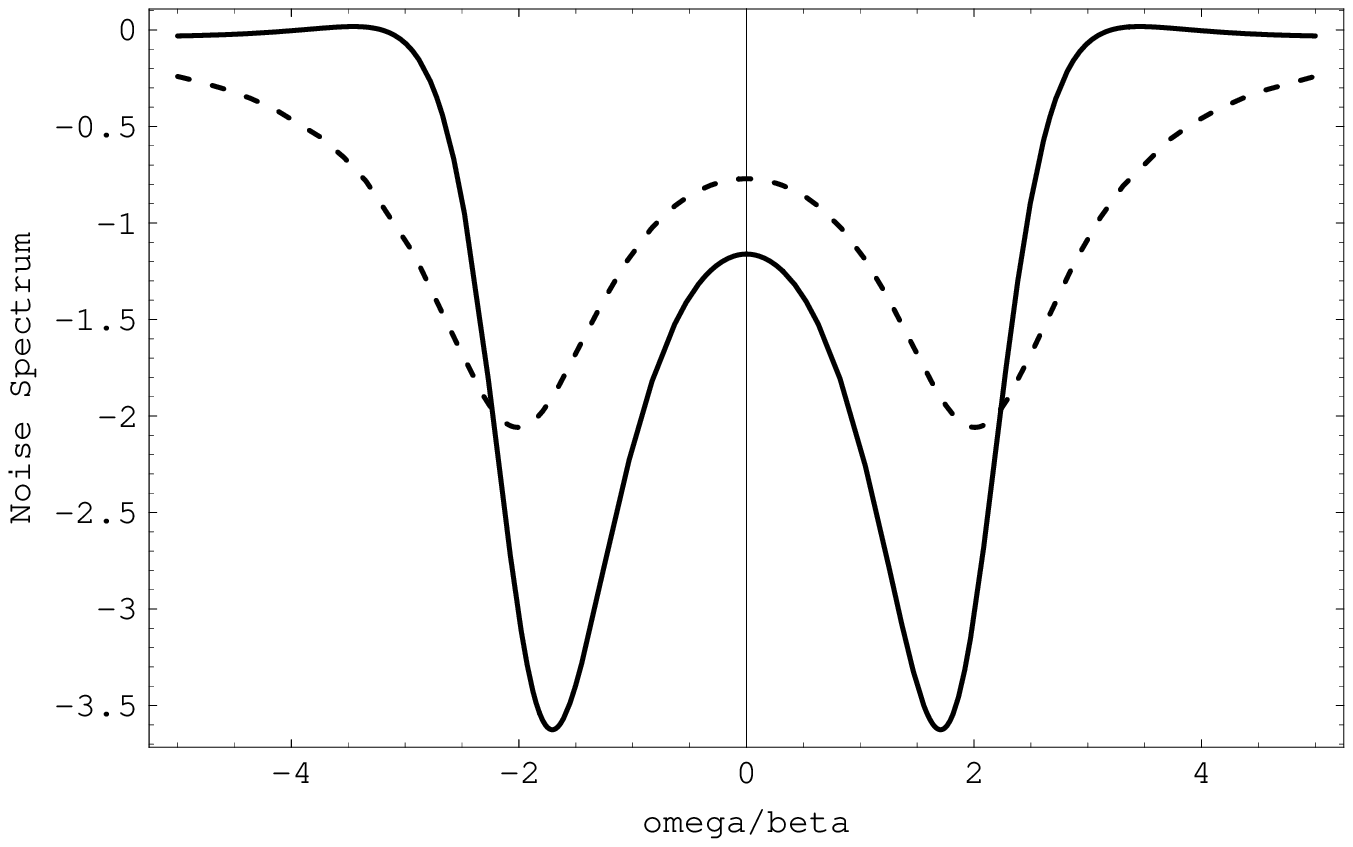}}
\end{figure}

\begin{figure}
\centerline{\psfig{figure=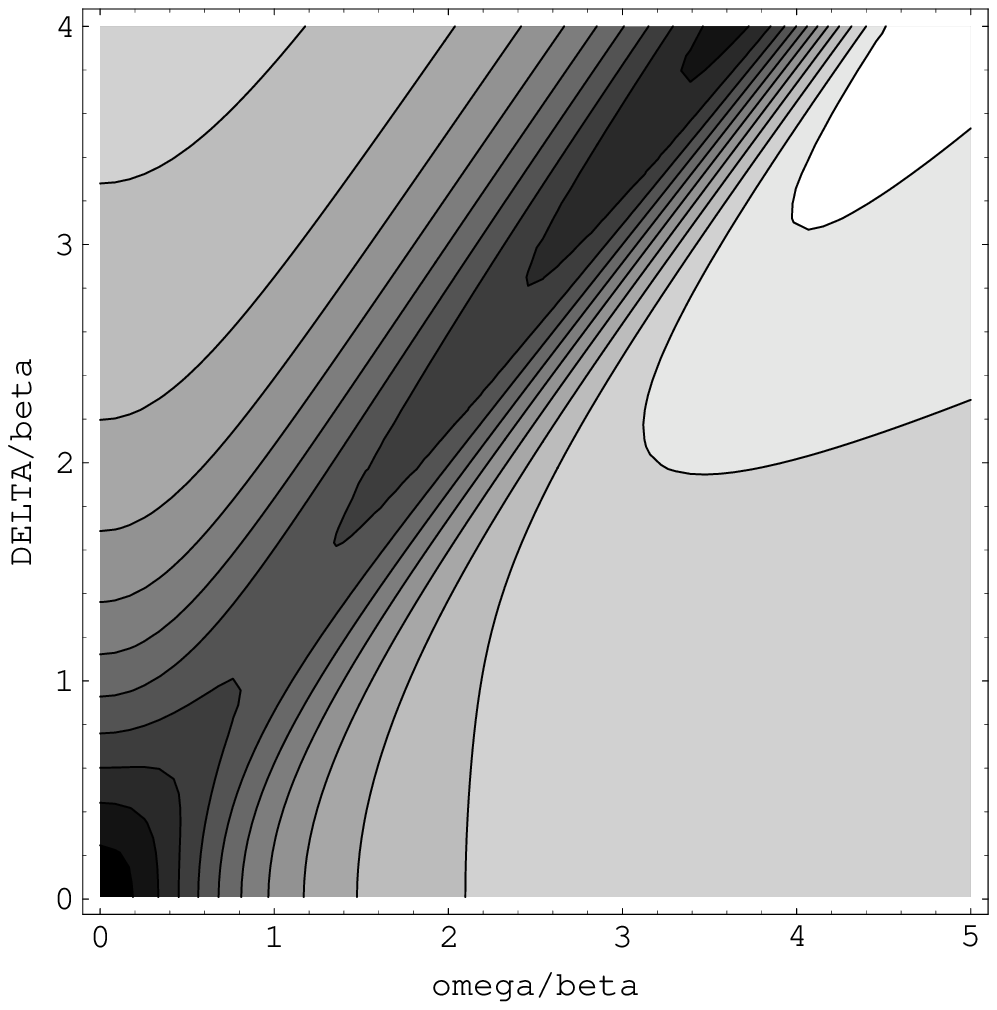}}
\end{figure}

\end{document}